\documentclass[a4paper,11pt]{article}
\usepackage{aaskaiid}
\usepackage{wrapfig}
\usepackage{tcolorbox}
\usepackage{orcidlink}

\def \deg         {\text{$^{\circ}$}}

\def \arcsec      {\text{$^{\prime\prime}$}}

\def \mujybeam    {$\mathrm{\mu}$Jy\,beam$^{-1}$}

\title{The Cosmic Ray Life Cycle in Galaxy Clusters}
\ShortTitle{The Cosmic Ray Life Cycle in Galaxy Clusters}

\author[1]{Francesco de Gasperin\orcidlink{0000-0003-4439-2627}}
\ShortName{F. De Gasperin et al.} 
\author[2]{Marcus Br\"uggen\orcidlink{0000-0002-3369-7735}}
\author[3]{Henrik Edler\orcidlink{0000-0002-4526-4806}}
\author[4]{Reinout~J. van Weeren\orcidlink{0000-0002-0587-1660}}
\author[1]{Tiziana Venturi\orcidlink{0000-0002-8476-6307}}
\author[1]{Gianfranco Brunetti\orcidlink{0000-0003-4195-8613}}

\affiliation[1]{INAF - Istituto di Radioastronomia, via P. Gobetti 101, Bologna, Italy}
\emailAdd{fdg@ira.inaf.it}
\affiliation[2]{Hamburger Sternwarte, University of Hamburg, Gojenbergsweg 112, D-21029, Hamburg, Germany}
\affiliation[3]{ASTRON, Netherlands Institute for Radio Astronomy, Oude Hoogeveensedijk 4, 7991 PD Dwingeloo, The Netherlands}
\affiliation[4]{Leiden Observatory, Leiden University, PO Box 9513, 2300 RA Leiden, The Netherlands}

\abstract{Within the cosmic web, gravitational energy, linked to the formation and growth of the Universe’s largest structures and the activity of active galactic nuclei (AGN), is converted to heat through processes such as turbulence and shock waves. These processes have a fundamental impact on the evolution of galaxy clusters. For example, they lead to the amplification of magnetic fields and the production of  cosmic ray (CR) electrons that emit continuum radio waves via synchrotron emission. This produces sources on scales of the entire hosting clusters.

These large radio sources in galaxy clusters are often classified based on their morphological appearance as radio halos, cluster radio shocks (radio relics), and other types. To understand the CR acceleration processes in galaxy clusters (and beyond), and to gain a comprehensive view of these sources, including their long-term interactions, SKA telescope should conduct both deep observations of a carefully selected sample of nearby clusters as well as shallower wide-area surveys.

Thanks to their capabilities — in particular the sensitivity to polarised and low-frequency emission — SKA-Mid (Bands 1 and 2) and SKA-Low are ideally suited to probing magnetic field structures in galaxy clusters, as well as the large reservoir of low-energy CRs that may be accelerated by yet-unexplored microphysical mechanisms. The high sensitivity to low-frequency emission will also be fundamental to detect the long term actions and interactions of these phenomena over gigayear timescales.}

\begin{document}
\maketitle
\newcommand{\actaa}{Acta Astron.} 
\newcommand{\araa}{Annu. Rev. Astron. Astrophys.} 
\newcommand{\aar}{Astron. Astrophys. Rev.} 
\newcommand{\ab}{Astrobiol.} 
\newcommand{\aj}{Astron. J.} 
\newcommand{\apj}{Astrophys. J.} 
\newcommand{\apjl}{Astrophys. J. Lett.} 
\newcommand{\apjs}{Astrophys. J. Suppl. Ser.} 
\newcommand{\ao}{Appl. Opt.} 
\newcommand{\apss}{Astrophys. Space Sci.} 
\newcommand{\aap}{Astron. Astrophys.} 
\newcommand{\aapr}{Astron. Astrophys. Rev.} 
\newcommand{\aaps}{Astron. Astrophys. Suppl.} 
\newcommand{\baas}{Bull. Am. Astron. Soc.} 
\newcommand{\caa}{Chinese Astron. Astrophys.} 
\newcommand{\cjaa}{Chinese J. Astron. Astrophys.} 
\newcommand{\cqg}{Class. Quantum Gravity} 
\newcommand{\gal}{Galaxies} 
\newcommand{\gca}{Geochim. Cosmochim. Acta} 
\newcommand{\icarus}{Icarus} 
\newcommand{\jcap}{J. Cosmol. Astropart. Phys.} 
\newcommand{\jgr}{J. Geophys. Res.} 
\newcommand{\jgrp}{J. Geophys. Res.: Planets} 
\newcommand{\jqsrt}{J. Quant. Spectrosc. Radiat. Transf.} 
\newcommand{\memsai}{Mem. Soc. Astron. Italiana} 
\newcommand{\mnras}{Mon. Not. R. Astron. Soc.} 
\newcommand{\nat}{Nature} 
\newcommand{\nastro}{Nat. Astron.} 
\newcommand{\ncomms}{Nat. Commun.} 
\newcommand{\nphys}{Nat. Phys.} 
\newcommand{\na}{New Astron.} 
\newcommand{\nar}{New Astron. Rev.} 
\newcommand{\physrep}{Phys. Rep.} 
\newcommand{\pra}{Phys. Rev. A} 
\newcommand{\prb}{Phys. Rev. B} 
\newcommand{\prc}{Phys. Rev. C} 
\newcommand{\prd}{Phys. Rev. D} 
\newcommand{\pre}{Phys. Rev. E} 
\newcommand{\prl}{Phys. Rev. Lett.} 
\newcommand{\psj}{Planet. Sci. J.} 
\newcommand{\planss}{Planet. Space Sci.} 
\newcommand{\pnas}{Proc. Natl Acad. Sci. USA} 
\newcommand{\procspie}{Proc. SPIE} 
\newcommand{\pasa}{Publ. Astron. Soc. Aust.} 
\newcommand{\pasj}{Publ. Astron. Soc. Jpn} 
\newcommand{\pasp}{Publ. Astron. Soc. Pac.} 
\newcommand{\rmxaa}{Rev. Mexicana Astron. Astrofis.} 
\newcommand{\sci}{Science} 
\newcommand{\sciadv}{Sci. Adv.} 
\newcommand{\solphys}{Sol. Phys.} 
\newcommand{\sovast}{Soviet Ast.} 
\newcommand{\ssr}{Space Sci. Rev.} 
\newcommand{\uni}{Universe} 

\section{Introduction}
Galaxy clusters increase their mass through the accretion of the Warm-Hot Intergalactic Medium (WHIM) from filaments and through mergers with groups (minor mergers) or other clusters (major mergers). Releasing up to $10^{64}$ ergs, mergers are the most energetic events in the Universe. This energy is dissipated through turbulence and low-Mach number shocks to microscopical scales, heating the intracluster medium \citep[ICM; for a review,][]{Brunetti2014}. There is now a good body of evidence showing that both mechanisms can accelerate particles to relativistic energies producing CRs.

A second relevant non-thermal component are magnetic fields.  $\mu$G magnetic fields appear to pervade galaxy clusters \citep[e.g.][]{Bonafede2010}, but essential properties of magnetic fields on all scales remain fairly unconstrained, such as their spatial spectra. Moreover, direct detections of magnetic fields in filaments and voids remain elusive. Current upper limits suggest field strengths of a few nG in filaments and as low as $10^{-16}$ G to $10^{-10}$ G in voids \citep{Vazza2021a, 2023MNRAS.523.6320H}. However, recent observations have revealed significant magnetic fields in the range $0.1-1$~$\mu$G in intercluster bridges, connecting pre-merging clusters \citep{Govoni2019}. Across all these environments, the plasma $\beta$ (thermal to magnetic pressure ratio) is found to be high, typically around 100 \citep{Carilli2002}.

\begin{figure}
    \centering
    \includegraphics[width=0.7\linewidth]{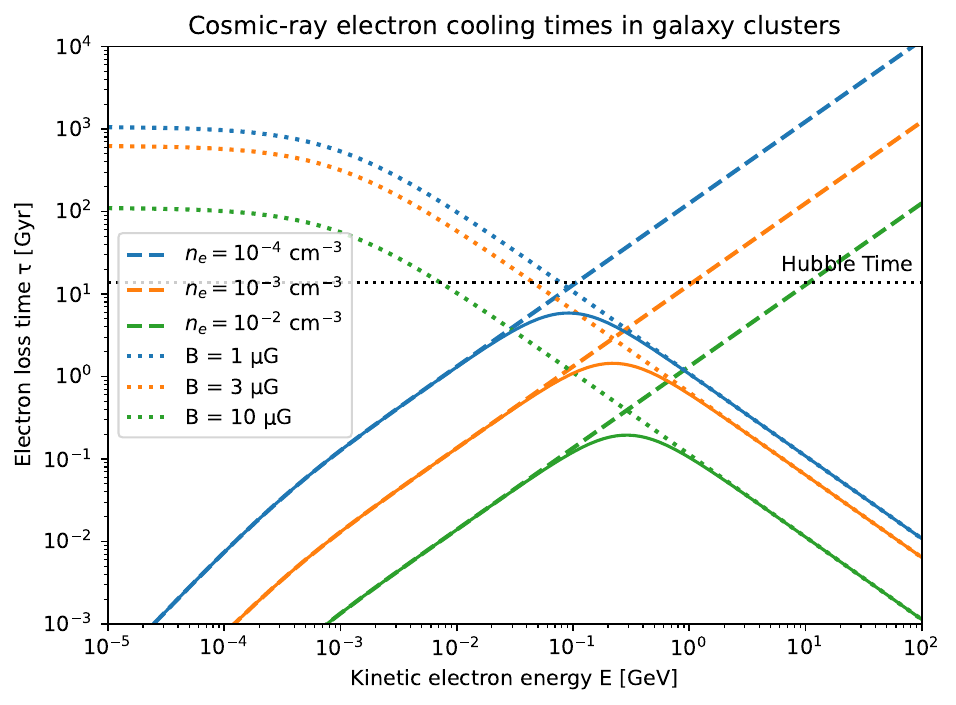}
    \caption{Cooling time of CR electrons under typical ICM conditions due to Coulomb and IC/synchrotron interaction as a function of their kinetic energy. The density and magnetic field strengths ranges from typical values in the outskirts of galaxy clusters to the central regions. CR electrons can remain long-lived if they are injected in low-density environments and at late cosmic times. Electrons with kinetic energy $E \simeq p\,m_{\mathrm{e}} c^{2} \simeq 10-100\,\mathrm{MeV}$ are expected to have the longest lifetimes in cluster outskirts $(n_{\mathrm{e}} < 10^{-4}\,\mathrm{cm}^{-3}$ and $B<1~\mu G)$, since this energy range maximizes Coulomb, synchrotron and inverse Compton (here derived at $z=0$) cooling times.}
    \label{fig:CRcooling}
\end{figure}

Due to the presence of magnetic fields, CRs lose energy via synchrotron radiation, steepening the emitted radio spectrum with time (see the SEDs in Fig.~\ref{fig:surveys}). CRs of energy higher than 10 GeV, that are responsible for GHz emission in a typical $\mu$G mgnetic field, have energy loss timescales of $\sim 10^8$ yr at $z = 0$. This losses increase at higher redshift due to inverse Compton. CRe of lower energies (100 MeV) instead, can survive for long periods of time in the typical conditions of cluster outskirts (see Fig.~\ref{fig:CRcooling}). These CR electrons can thus form a pool of fossil electrons that may be key to explaining the phenomenology of diffuse radio sources in galaxy clusters. Synchrotron-dark CRs are therefore expected to accumulate in the ICM, but their observation has remained elusive \citep[e.g.][]{Pinzke2013}. 

\subsection{Main CR acceleration mechanisms}

Collisionless \textit{shock waves} are commonly generated in the ICM through cluster mergers and matter accretion. A variety of shock types arise during structure formation such as: 1. \textbf{internal shocks}, produced by bulk flows, minor mergers, or turbulent motions within the ICM \citep[e.g.][]{Pfrommer2006}; 2. \textbf{accretion shocks}, marking the transition between the cosmic web and the cluster environment where cold, low-density gas first encounters the hot, virialized ICM \citep[e.g.][]{Vazza2009}; 3. \textbf{merger shocks}, produced by supersonic collision of subcluster gas halos, they propagate outwards through the ICM after core passage \citep[e.g.][]{vanWeeren2010a}; 4. \textbf{runaway shocks} late development of merger shocks. In a uniform medium shocks would weaken with time, however in clusters the shock moves down a steep density gradient that helps the shock to maintain or even increase its strength \citep[e.g.][]{Zhang2019}; and 5. \textbf{equatorial shocks} located at the equatorial plane perpendicular to large-scale filaments accreting matter into a cluster \citep[e.g.][]{Akamatsu2017a}. Most of these shocks have low Mach numbers (typically a few), except for accretion shocks in the cluster outskirts, where cold gas from voids is accreted and Mach numbers can reach $\sim100$. However, these external shocks occur in low-density regions and contribute little to the ICM's energy budget. As a result, internal and merger shocks are the primary drivers of ICM heating. 

\textit{Turbulence} is a well-established feature of the ICM, with high Reynolds numbers $R_e \gtrsim 1000$ indicating low viscosities. 
On megaparsec scales, turbulence is primarily injected by large-scale processes such as cluster mergers and matter accretion. In contrast, on smaller (tens of kiloparsecs) scales, turbulence is fuelled by AGN, galaxy dynamics, and thermal instabilities.

Recent observations from the XRISM satellite have provided compelling evidence of turbulent motions within the ICM \citep{2025arXiv251012782Z}. High-resolution X-ray spectroscopy from XRISM’s Resolve instrument has enabled direct measurements of line broadening in heavy-ion emission, revealing velocity dispersions on the order of a few hundred kilometers per second \citep{2025PASJ...77S.270F}. These findings indicate highly subsonic turbulence, with turbulent heating supplying roughly 40\% of the local radiative cooling rate. XRISM data of the Coma cluster appear to be inconsistent with a Kolmogorov spectrum in the case of homogeneous and volume filling turbulence \citep{XrismCollaboration2025}. Subsequent analysis have removed this tension showing a consistency between simulated clusters and XRISM data due to the fact that turbulence in simulated clusters is not homogeneous \citep{Vazza2025}. These measurements confirm that turbulence contributes substantially to the non-thermal pressure support within clusters, complementing thermal and bulk flow components. 

\section{The cosmic ray life-cycle in galaxy clusters}

There are a number of phenomena linked to the (re-)acceleration of CRs within a cluster environment on a scale beyond the individual galaxies. We will now discuss the known phenomena from an observational perspective.

\begin{wrapfigure}[17]{r}{0.45\textwidth}
\vspace{-1cm}
\centering
\includegraphics[width=0.35\textwidth]{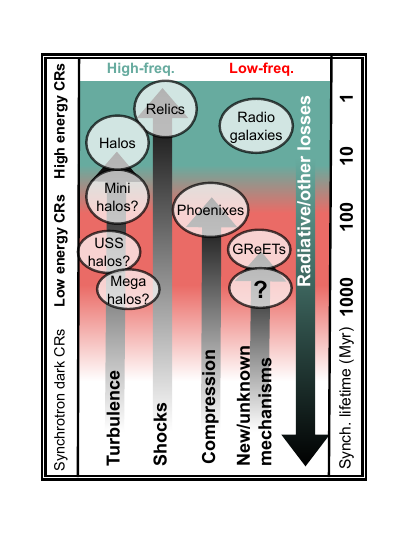}
\caption{CRs (re)-energisation/ageing processes (arrows) in clusters and related observational phenomena (ovals). CR energy increases from bottom to top.}\label{fig:CR}
\end{wrapfigure}

Low-frequency observations trace CR energised by merger shock waves, producing \textit{radio relics} \citep{Ensslin1998, vanWeeren2010a}, Mpc-size arc-shaped sources at the periphery of galaxy clusters. The luminosity of radio relics is proportional to the mass of the hosting cluster, effectively linking the phenomena to the dissipation of mergers' energy \citep[see e.g.][]{deGasperin2014c, Jones2023}. Observations also confirmed that radio relics are likely generated by series of shocks with non-uniform properties \citep{deGasperin2021}, as predicted by simulations \citep{Skillman2013, Inchingolo2022}.

In \cite{Brunetti2001} and \cite{Petrosian2001}, the authors proposed that regions of turbulence in the ICM can re-accelerate CRs and produce \textit{radio halos}, centrally located low-surface brightness sources that follow the ICM distribution. In line with the ``turbulence'' theory it has been proved that: radio halos are associated to merging clusters \citep{Cassano2010, Cuciti2021a}, their occurrence increases with the cluster mass \citep{Cuciti2015} and the spectra of halos in less energetic merges is steep \citep{Pasini2024}. In the same region, a certain amount of CR electrons is also injected by the decay of charged pions produced by proton-proton collisions \citep{Dolag2000}.

Conversely, \textit{radio mini-halos} can be found in relaxed cool-core systems. Their size is a few 100 kpc and their radio emission surrounds the central AGN. It has been proposed that radio mini-halos are either of hadronic origin or produced by turbulent (re-)acceleration.
In the second case, the turbulence is induced by gas sloshing that induce large-scale turbulent motions while keeping the relaxed cool core intact \citep[e.g.][]{Riseley2022}. Minor mergers and/or off-axis mergers could be a source of such turbulence. \citet{Savini2018} have found low-frequency faint halo-like emission beyond the cold fronts in RX J1720.1+2638. This halos might be due to turbulence injected by a minor merger, that did not disrupt the cool-core. This implies that mini-halos and radio halos could potentially co-exist. 

In the centre of relaxed clusters, gas is expected to cool efficiently via thermal Bremsstrahlung reducing its pressure support and letting more gas from the surrounding regions flow inwards \citep[cooling flow;][]{Fabian2012, McNamara2012}. However, this large amount of cold gas is not detected with X-ray observations \citep{Kaastra2001, Peterson2001}. Feedback from \textit{AGN} is thought to be the main source of energy to counterbalance these radiative losses. AGN feedback energy can be deposited into the ICM through bubbles that rise buoyantly to larger cluster radii, as well as through turbulence, weak shocks, and sound waves generated by AGN outbursts \citep{Forman2007, Zhuravleva2014}. This action is witnessed by the presence of large cavities in the X-ray emitting ICM filled with radio emitting CRs. Plasma in older cavities lost all energetic CRs and can be detected only at low-frequencies. This plasma can be a reservoir for a number of CR re-acceleration processes. Empirical correlations can be used to infer the mechanical energy output of AGN from radio observations \citep{Birzan2004}. The amount of energy output and its redshift evolution, are crucial to constrain models of feedback \citep[e.g.][]{Gaspari2011}. AGN feedback is thought to be also responsible to set the upper limit to the masses of galaxies, halting star formation, and determining the observed properties of early-type galaxies \citep{DeLucia2007}. 

\begin{wrapfigure}[17]{r}{0.47\textwidth}
\centering\vspace{-.5cm}
\includegraphics[width=\linewidth]{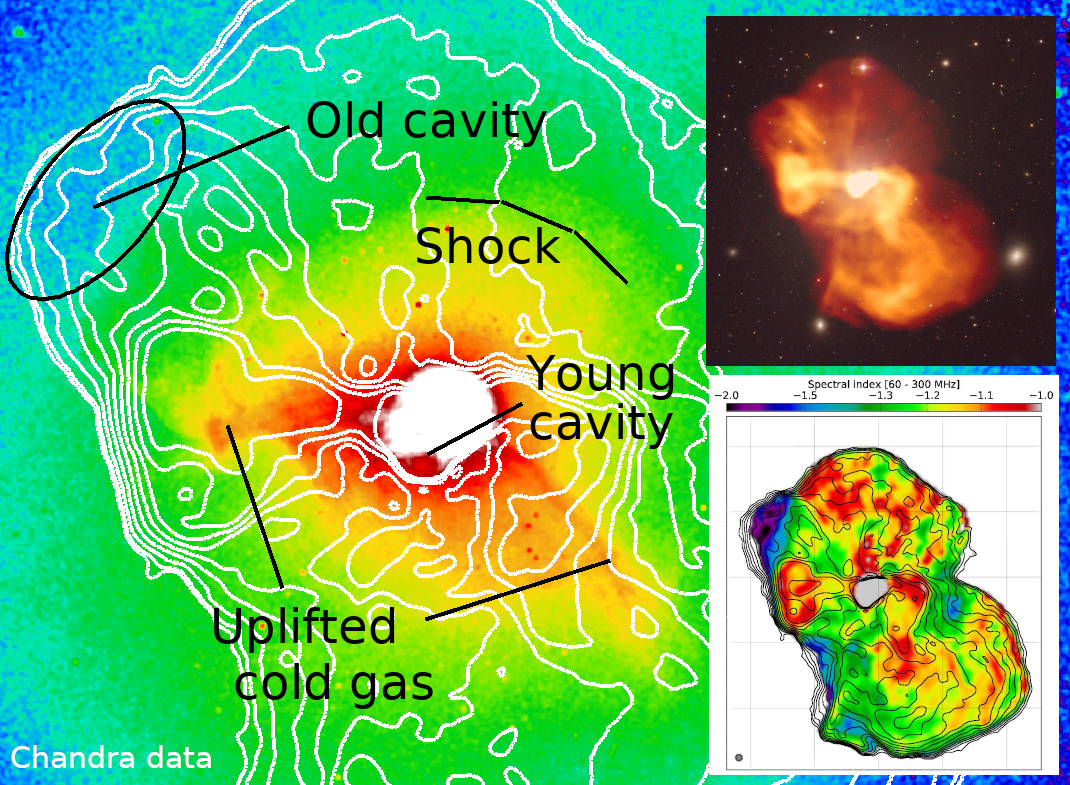}
\caption{M87 is one of the best example of AGN feedback in action. Here we labelled some of the features present in X-ray data. In the small panels, the LOFAR 54 MHz image and the spectral index map showing active/young regions (red) and dead/old regions (blue).} \label{fig:m87}
\end{wrapfigure}

\textit{Star-formation} (SF) in galaxies is another primary CR source in galaxy clusters. These electrons can then be transported outside of the galaxies and mixed with the ICM via dispersive and advective transport processes. The latter, in the form of ram pressure stripping, leads to extended non-thermal tails behind SF-galaxies \citep{skaignesti}. Similar to AGN, star-forming galaxies also contribute to the general low-energy CR population in the cluster, that is available as seed electrons for re-acceleration by shocks and turbulence. 

Low-frequency observations are unveiling a further vast phenomenology of re-energised plasma (see Fig.~\ref{fig:CR}), showing that CR life-cycle in galaxy clusters is far more complex than the simple picture: acceleration $\rightarrow$ fading. Plasma, in most cases of AGN origin, can be re-energised though a variety of mechanisms including compression \citep[radio phoenixes;][]{Ensslin2001}, shocks \citep[some radio relics;][]{vanWeeren2017}, turbulence \citep[mini halos;][]{ZuHone2013} or complex plasma interactions \citep[Gentle Re-Energised Tails - GReETs;][]{deGasperin2017}. These inefficient mechanisms produce sources with extremely steep spectra invisible to radio telescopes observing at $>1$~GHz. Within this variety of theoretically proposed but observationally poorly studied mechanisms, resides the possibility to expand our knowledge on the micro-physics of the ICM thanks to the advent of new instruments such as SKA.

\begin{figure}[b]
\centering
\includegraphics[width=.49\textwidth]{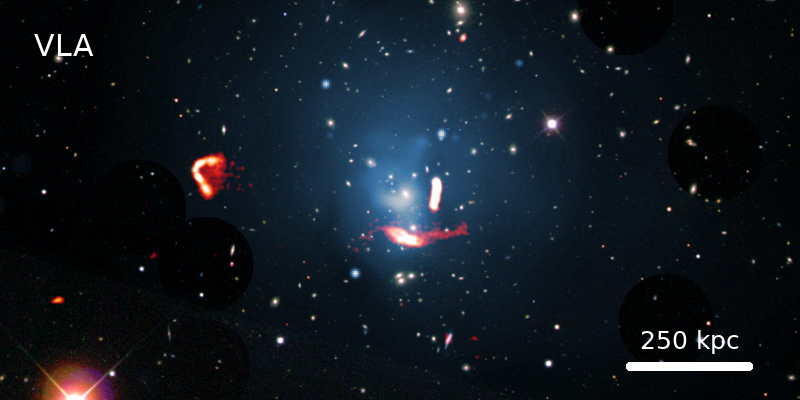}
\includegraphics[width=.49\textwidth]{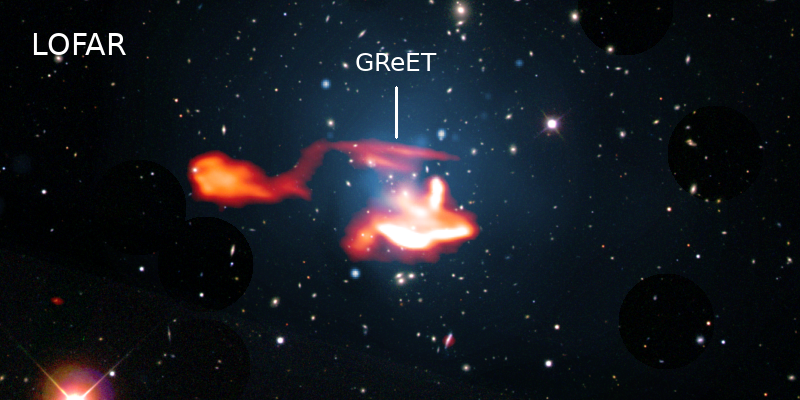}
\caption{\textit{Left:} the galaxy cluster Abell 1033 at conventional frequencies (1.4 GHz). \textit{Right:} the discovery of the first GReET, a new type of radio source visible uniquely at very low frequencies.} \label{fig:greet}
\end{figure}


\section{Cluster boundaries and beyond}

The next grand step in the field will come from the exploration of cluster boundaries and cosmic filaments. About half of the baryons of the Universe resides in the filaments of the large scale structure under the form of a warm ($10^{6-7}$~K) and rarefied ($\lesssim 10^{-4}$ particles cm$^{-3}$) gas, the so-called warm-hot intergalactic medium \citep[WHIM;][]{Martizzi2019}. Due to their low emissivity, our observational knowledge of WHIM distribution and condition is currently very limited. A number of close (few Mpc) pre-merger pairs of galaxy clusters showed the presence of X-ray \citep{Sakelliou2004, Akamatsu2017a}, SZ \citep{PlanckCollaboration2015b}, and recently discovered radio emission \citep[radio bridges, see below;][]{Govoni2019, Botteon2020b}. However, in these special regions the interaction between the two merging clusters already altered the WHIM conditions e.g. through the injection of energy via shocks \citep{Gu2019} or the development of widespread turbulence \citep{Brunetti2020}. 
\begin{wrapfigure}[21]{r}{0.55\textwidth}
\centering
\includegraphics[width=0.55\textwidth]{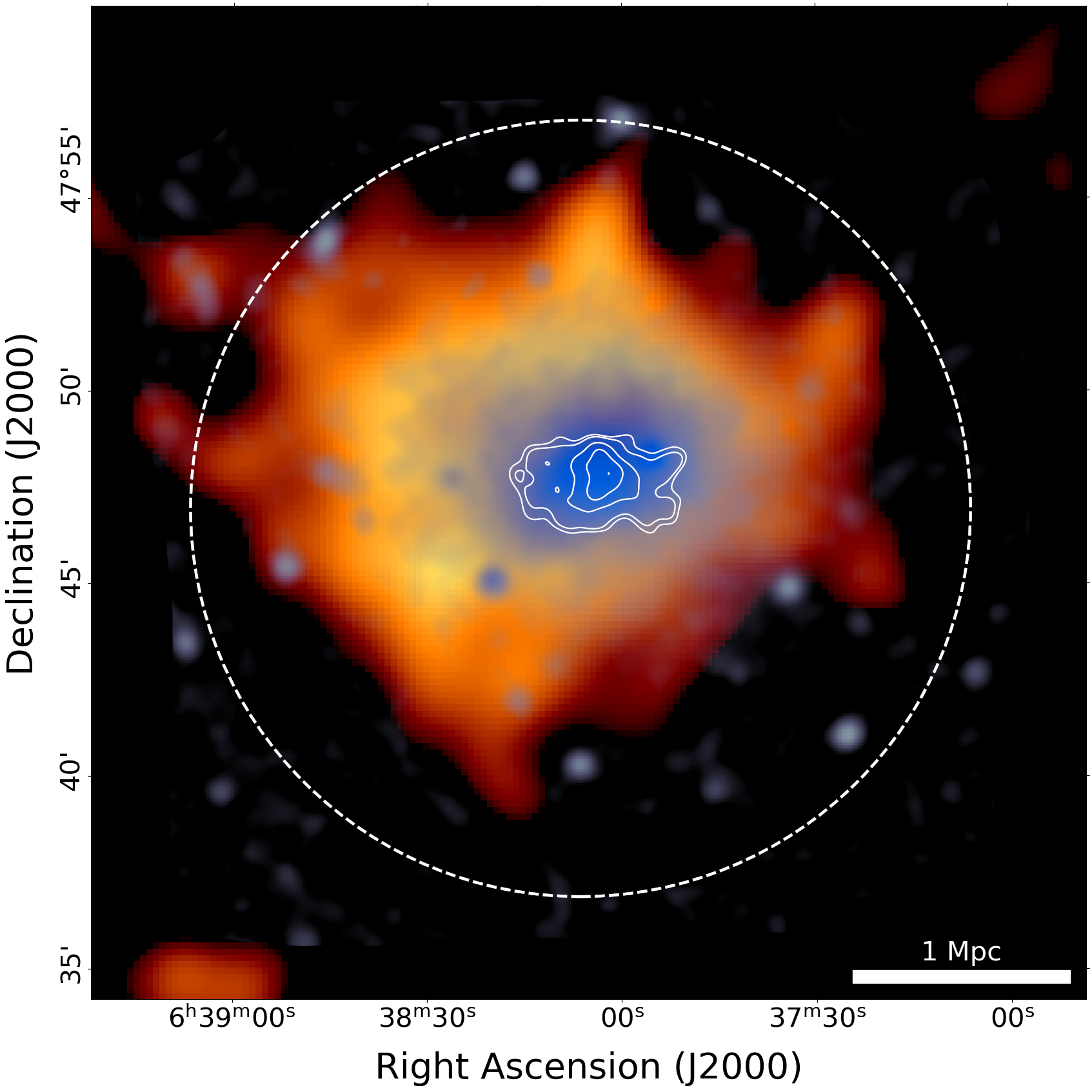}
\caption{In orange the first detection of a radio Megahalo \citep{Cuciti2022}, with a size of 3 Mpc it extends up to the virial radius (dotted line), occupying a volume 30 times larger than classical radio halos (outlined with contours). In blue the X-ray emission from the ICM.}\label{fig:megahalo}
\end{wrapfigure}
Direct detections of WHIM in longer filaments were so far limited to the enhanced X-ray emission between Abell 3667 and Abell 3651 (13 Mpc projected), discovered observing in the soft X-rays with eROSITA \citep{Dietl2024}. Filaments material located close to the interface with a galaxy clusters are the easiest target to foster this exploration, but even those have been rarely accessible.

The first step towards the detection of cosmic filaments in the radio band came from \citet{Govoni2019} and \citet{Botteon2020}, that presented the first detections of radio \textit{bridges} connecting pre-merger pairs of massive clusters. These observations demonstrate that relativistic electrons and magnetic fields can drain energy generated on scales larger than cluster-scales, and potentially probe the dynamics of large-scale structures and the dissipation energy mechanisms therein.

Both clusters and filaments are expected to be surrounded by accretion shocks where the gas is heated \citep{Sunyaev1972}. However, radio emission from the accretion-shock, or other clear detections, have remained elusive. One step forward in the exploration of cluster boundaries came with the discovery of a new population of very large ($\sim 3$~Mpc) radio sources filling the entire cluster volume (Fig.~\ref{fig:megahalo}). These \textit{Megahalos} trace the presence of magnetic field and CR far into the cluster outskirts, where direct observations of the ICM will be possible only with the next generation of X-ray satellites \citep{Walker2022}. These sources provide unique insights on clusters outskirts conditions, such that the energy associated with non thermal components (CR and magnetic fields) increases compared to the energy in the thermal counterpart \citep{Cuciti2022}.

Radio emission in filaments is linked to the seeding of primordial magnetic fields (\textit{magnetogensis}), one of the big mysteries of modern astrophysics. MHD simulations show that due to the long dynamical time scale of filaments, the associated magnetic field strength is closely related to the \textit{seed} cosmic field \citep{Vazza2014}. Furthermore, the spatial distribution of the radio emission can distinguish between the two scenarios: (1) a relevant seed field present before the CMB formation, (2) a seed field injected by AGN jets and galactic magnetised outflows \citep{Donnert2009}.

\begin{figure}
\centering
\includegraphics[width=0.7\textwidth]{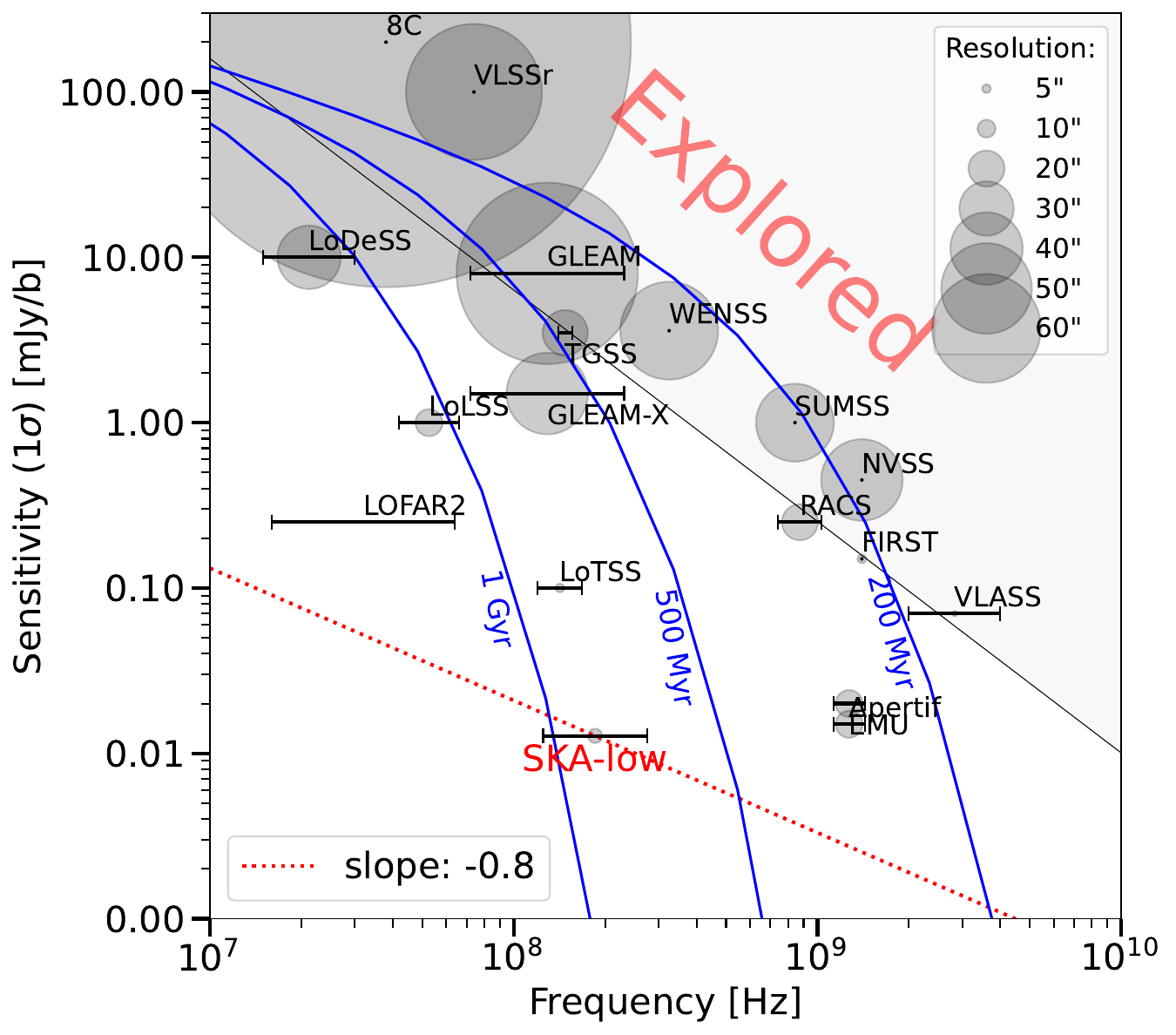}
\caption{Location in a sensitivity-frequency plot of a possible SKA-low all-sky survey compared to past and on-going large area surveys. Blue lines are the SED of a radio source after ageing \citep[following a standard JP model;][]{Jaffe1973a} for 0.2, 0.5, and 1 Gyr.}\label{fig:surveys}
\end{figure}

\section{SKA advancements in the field}

According to the theory of diffusive shock acceleration \citep[DSA;][]{Bell1978}, the low-Mach number of cluster shocks requires an unrealistic fraction of shock energy to be channelled in the acceleration of CR electrons from the thermal pool \citep[e.g.][]{Vazza2014, Botteon2020}. To alleviate the problem, it has been proposed that shocks can re-accelerate aged electrons originally energised by radio-galaxies \citep{Markevitch2005, Kang2011, Pinzke2013}. There has been some observational support for this scenario \citep{vanWeeren2017}. Furthermore, low-frequency spectra show that the fading of the accelerated particles in the post-shock region is slower than predicted \citep{deGasperin2020a}, possibly implying some unexpected reacceleration mechanisms, such as turbulence, downstream of the shock front \citep{Kang2017}. It is fair to say that both acceleration and fading mechanisms in radio relics is still unclear and very much linked to the fate of aged CRs of AGN origin that only deep low-frequency radio observations can explore. High-resolution and high-sensitivity observations of these sources are required to make any advancements. Crucial will be the combined sensitivity of the SKA-Low in detecting old CRe populations, shock precursors and high resolution spectra of the fading CRe. In the shock region, bright filaments and downstream filaments are likely magnetically dominated. So far, the precise study of their polarisation properties has only been done in a handful of favourable cases \citep[e.g.][]{deGasperin2022, DiGennaro2021, Rajpurohit2022}. Thus, the capabilities of SKA-Mid will have a profound impact on the study of relics.

For radio halos, the turbulent (re)acceleration model predicts that their radio spectrum is steep at the time of formation and subsequently decays \citep{Donnert2013}. Similarly steep spectra are expected also in less energetic mergers \citep{Brunetti2008}. A remarkable consequence is the prediction of a large population of ultra-steep spectrum (USS) radio halos that are expected to glow at low radio frequencies \citep{Cassano2006}. In alignment with expectations, a few USS radio halos have already been found \citep[e.g.][and references therein]{Pasini2024, Magolego2025}, although the bulk of the population is still missing. Sensitive, large-scale surveys at low-frequency covering a large number of clusters are the next step to validate predictions. To obtain complete samples of clusters, the SKA will provide synergies with SZ-telescopes that operate in the southern hemisphere \citep[e.g.][]{Hilton2021}. Finally, SKA-Mid (band 1 and 2) will have the sensitivity to detect radio halo emission at high angular resolution, thus reducing beam depolarisation, and detect the internal magnetic structure of radio halos \citep{Govoni2013}.

Radio halos and relics produce CRs energetic enough to be detectable at GHz frequencies, however they represent just the tip of the iceberg. A silent majority of sources powered by less-energetic CRs is only now becoming visible thanks to sensitive low-frequency observations. A striking example are GReETs \citep[Fig.~\ref{fig:greet};][]{deGasperin2017}, where aged populations of CR electrons from a radio galaxy are maintained synchrotron-bright by a continuous, gentle interaction with the perturbed ICM. The spectral index\footnote{$F \propto \nu^\alpha$ where F is the radio flux, $\nu$ the frequency, and $\alpha$ the spectral index.} of this source is $\alpha\simeq-4$, which makes it invisible at frequencies above just a few hundreds MHz. The number of GReETs is unknown and other cases of low-efficiency CR acceleration phenomena may still be undetected. Advancements will come from the high sensitivity of SKA-Low (detection of more examples) and from the possibility to study the polarisation properties of some of these sources with deep observations of SKA-Mid.

In \cite{2024Galax..12...19V}, the authors show how radio galaxies in clusters serve as dominant sources of magnetic fields and cosmic-ray electrons, which are crucial for diffuse nonthermal radio emissions such as halos, relics, and mini‑halos. Quantitative models show that a typical population of cluster radio galaxies can inject enough CR electrons to account for the observed radio halo powers but only supply around 30\% of the area in cluster outskirts even after 1 Gyr of mixing. Numerical simulations reveal turbulent gas motions and weak shocks, with velocities dispersion of 300~km/s, are essential for redistributing and re-accelerating these fossil electrons to sustain observable radio features over hundreds of Myr and Mpc scales. Moreover, it has been found that re-acceleration processes are crucial to maintain a volume-filling reservoir of fossil electrons for several Gyrs after the first injection by radio jets. This is essential for explaining cluster-wide emission and other radio phenomena in galaxy clusters. A first direct observation of this synchrotron-dark population by SKA can come from older bubbles form large, peripheral cavities excavated by the AGN action that are often radio silent (``ghost cavities''). The most energetic electrons inside these cavities lost most of their energy, but the low-energetic electrons still shine at the lowest radio frequencies \citep[see spectral map in Fig.~\ref{fig:m87} and e.g. ][]{Giacintucci2020}. Therefore, both the study of AGN duty cycle and integrated feedback output require low-frequency data. However, studies of cavity systems at low-frequencies are limited by the poor sensitivity and angular resolution of the instruments. These limitations are exacerbated at higher redshifts and little is known on the evolution of radio-feedback through cosmic time. 

Finally, a crucial part of the non-thermal components in galaxy clusters are magnetic fields. The regions where they are strong can be detected with high-frequency radio observations relatively easily, however the vast majority of cluster volume is filled with weak magnetic fields that can be revealed only with low-frequencies synchrotron radiation or with advanced techniques such as rotation measure studies \citep[e.g.][]{Bonafede2010}.

At the cluster periphery, accretion shock should be strong enough (Mach number $\gtrsim 10$) to accelerate CR in the WHIM \citep{Miniati2000, Keshet2004}. A few attempts were made in recent years, resulting either in non-detections or false positives \citep{vanWeeren2009b, Bagchi2002, Hodgson2021a}. With SKA-Mid we might be able to reach the necessary sensitivity to detect the strongest portion of these shocks, the main uncertainty being the local magnetic field. Their polarisation and spectral properties will enable a better identification of these sources. In general, polarised diffuse emission in and around clusters has the chances to become a relevant science case if the contamination from the galactic foreground can be properly taken into account.

Moving outside clusters, current radio facilities, such as LOFAR, MeerKAT, and ASKAP, have only attempted a statistical detection of radio emission from filaments \citep{Vernstrom2021}, this because their synchrotron radiation is extremely faint, well below the sensitivity limits. The unparalleled sensitivity of SKA will be transformative, enabling the possibility of routine detection of filament emission if conditions (magnetic field strengths and shock acceleration efficiencies) will turn out to be favourable or put the strongest constrains on these parameters in case of non detection \citep{skacuciti, Vazza2015, Vacca2024}. In addition, SKA's broad frequency coverage and polarimetric capabilities will allow detailed studies of the magnetic properties of filaments. High-fidelity polarization maps will measure Faraday rotation and depolarization effects, providing direct constraints on magnetic field strength, coherence length, and topology.

\section{A SKA-Low survey to study the cosmic web}
\label{sec:survey}

\begin{figure}
    \centering
    \includegraphics[width=1\linewidth]{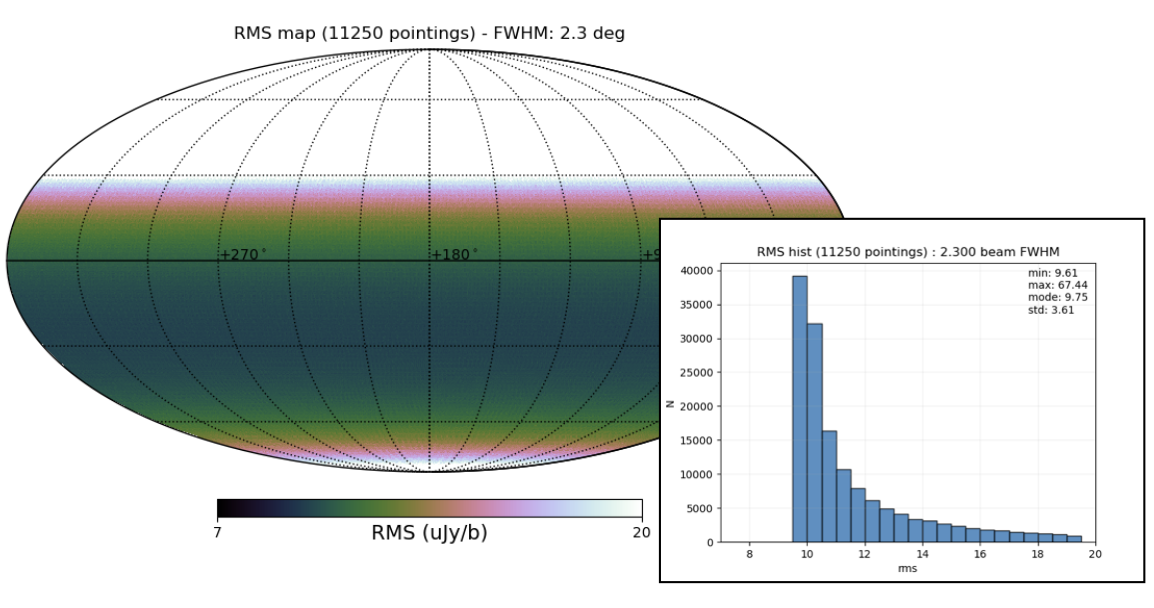}
    \caption{RMS noise map of the simulated SKA-Low survey described in Sec.~\ref{sec:survey}. The declination gradient is due to the dipole beam combined with a uniform time integration. The histogram shows the number of pixels per bin of rms noise.}
    \label{fig:surveymap}
\end{figure}

The large angular size of clusters outskirts and cosmic filaments combined with the (expected) low strength of the magnetic fields, makes the detection of synchrotron radiation in these regions rather difficult. However, the sensitivity of SKA-Low will make a difference, at least in the study of selected nearby clusters and known filaments. The main limitation, according to the specification of SKA-Low AA4, is the resolution that leads to confusion-limited observations rather quickly (few observing hrs) unless the longest baselines are up-weighted towards uniform weighting. An important difference will also come from the selected bandwidth, with higher frequencies needed to reduce the confusion noise, while lower frequencies are needed to detect steep spectrum sources.

The best way to collect data with the SKA-Low is through a large, deep survey as this would maximise the observing efficiency and streamline the data processing. We simulated a possible survey of this type using 2 simultaneous beams (150 MHz band each, covering $125-275$~MHz) up to declination $+30\deg$. The final products will have a resolution of about 8\arcsec (robust 0) and a sensitivity of 10~\mujybeam{} (including confusion, derived using the SKA Sensitivity Calculator\footnote{\url{https://sensitivity-calculator.skao.int/}}). The footprint would be covered in 5625 hrs (1 hr per pointing) going about an order of magnitude deeper than the LOFAR Two Meter Sky Survey, see Fig.~\ref{fig:surveys} \citep{Shimwell2017}. A uniform sensitivity coverage would require a gradual increase of the integration time for pointings away from Dec~$=-30\deg$. The flexibility of SKA-Low in AA4 (e.g. the use of sub-arrays), will help in optimising the scheduling of the survey.

With such a survey we would be able to detect about 4000 radio halos in galaxy clusters, up to redshift $z=0.6$ \citep{skacassano}. Furthermore, the survey would enable the detection of radio relics in high-redshift cluster and the missing population of faint relics. At the same time, assuming that the surface brightness of megahalos increases with the mass of clusters \citep[as suggested by][]{Cuciti2022}, such a survey would allow us to potentially unveil megahalos in almost all galaxy clusters  of the Planck sample up to $z\sim 0.6$. According to the cosmological simulations presented in \cite{skacuciti}, with the proposed observations we will also be able to detect at least the brightest parts of the faint emission from shocked regions of cosmic filaments.
While such a survey would be transformative to study the CR life-cycle in clusters, the main limitation will be the resolution required to morphologically disentangle sources as well as study the details of the internal source structure.

The AA4 configuration of SKA-Low will reach a maximum baseline length of 74 km with a thick baseline coverage. Compared to AA* this will enhance the survey speed. The image resolution will remain comparable to GMRT at 610 MHz and LOFAR at 150 MHz. The improved $uv$-coverage will enable a faster convergence in self-calibration, higher dynamic range and higher image fidelity.


\bibliographystyle{abbrvnat-maxbibnames4}
\bibliography{ref,lit2}

\end{document}